\documentclass[aps,prl,twocolumn,groupedaddress,showpacs]{revtex4}

\usepackage{graphicx}
\usepackage{bm}
\usepackage{amsmath}

\begin{document}


\title{Vortex molecules in coherently coupled two-component Bose-Einstein condensates}


\author{Kenichi Kasamatsu$^1$}
\author{Makoto Tsubota$^1$}
\author{Masahito Ueda$^2$}

\affiliation{
$^1$Department of Physics,
Osaka City University, Sumiyoshi-Ku, Osaka 558-8585, Japan 
\\
$^2$Department of Physics, Tokyo Institute of Technology,  
Meguro-ku, Tokyo 152-8551, Japan}


\date{\today}

\begin{abstract}
A vortex molecule is predicted in rotating two-component Bose-Einstein condensates whose internal hyperfine states are coupled coherently by an external field. A vortex in one component and that in the other are connected by a domain wall of the relative phase, constituting a ``vortex molecule", which features a nonaxisymmetric (pseudo)spin texture with a pair of {\it merons}. The binding mechanism of the vortex molecule is discussed based on a generalized nonlinear sigma model and a variational ansatz. The anisotropy of vortex molecules is caused by the difference in the scattering lengths, yielding a distorted vortex-molecule lattice in fast rotating condensates. 
\end{abstract}

\pacs{03.75.Lm, 03.75.Mn, 05.30.Jp}

\maketitle

Topological defects appear in cross-disciplinary subfields of physics as long-lived excitations, constrained by the topology of the order parameter \cite{Pisman}. A prime example is quantized vortices which play a key role in understanding of superfluidity \cite{Donnely}. When a system has a multicomponent order parameter, it is possible to excite various exotic topological defects which have no analogue in systems with a single-component order parameter. 

An atomic-gas Bose-Einstein condensate (BEC) offers an ideal testing ground to investigate such topological defects, because almost all parameters of the system can be controlled to the extent that state engineering is possible. Because alkali atoms have a spin degree of freedom, multicomponent BECs can be realized if more than one hyperfine spin state is simultaneously populated \cite{Hall,Stenger}. A quantized vortex and a vortex lattice in BECs have been created experimentally by several techniques \cite{Matthews,Leanhardt,Madison,Haljan}. The methods reported in Ref. \cite{Matthews,Leanhardt} utilized internal degrees of freedom of BECs, creating unconventional vortices described by multicomponent order parameters. The structure of single vortex states in systems with multicomponent order parameters was investigated in Refs. \cite{Ho,Yip,Mueller}. In addition, it has been predicted theoretically that fast rotating two-component BECs exhibit a rich variety of unconventional vortex structures \cite{Mueller,Mueller2,Kasamatsu}.

In this Letter, we study the vortex structure of rotating two-component BECs whose internal states are coupled coherently by an external driving field. This coupling can be achieved experimentally as reported in Refs. \cite{Hall2,Matthews2}, where Rabi oscillations between the two components were observed. If the strength of the coupling drive is increased gradually from zero and its frequency is gradually ramped to resonance, one can obtain a stationary state with a nearly equal-weight superposition of the two states \cite{Matthews2}. Here, we study stationary states of two-component BECs with an external rotation as well as internal coupling. Combination of these two effects enables us to explore a new regime of rich vortex structures beyond the conventional binary system \cite{Ho,Mueller2,Kasamatsu,Ripoll3}; the two components interact not only through their mean-field interactions but also through the relative phase of the order parameters. We find that such two-component BECs exhibit unique vortex structures; {\it a vortex in one component and that in the other form a stable vortex-antivortex pair} \cite{tyuu2}, {\it i.e., a ``vortex molecule", bound by a domain wall in the space of relative phase}. The psudospin representation of the molecule reveals a nonaxisymmetric spin texture with a pair of {\it merons} \cite{Seppala,Moon,Steele}. The internal coupling controls the binding force of the vortex molecule and affects significantly the lattice structure of vortices in fast rotating two-component BECs. 

We consider two-component BECs of atoms with mass $m$ residing in two hyperfine states in a rotating frame with rotation frequency ${\bf \Omega} = \Omega {\bf \hat{z}}$. The BECs are described by the condensate wave functions $\Psi_{i}$ ($i=1,2$). The equilibrium state of the system is obtained by minimizing the energy functional
\begin{eqnarray}
E[\Psi_{1},\Psi_{2}] = \int d {\bf r} \biggl\{ \sum_{i=1,2} \biggl( \Psi_{i}^{\ast}  h_{i} \Psi_{i} 
+ \frac{u_{i}}{2} |\Psi_{i}|^{4}  \biggr)  \nonumber \\
+ u_{12} |\Psi_{1}|^{2} |\Psi_{2}|^{2} 
-\omega_{\rm R} (\Psi_{1}^{\ast} \Psi_{2} e^{i\Delta t} + \Psi_{1} \Psi_{2}^{\ast}e^{-i\Delta t}) \biggr\}.
\label{coupleGPfunc} 
\end{eqnarray}
In this work, we confine ourselves to the two-dimensional problem. Here, we measure the length and energy in units of $b_{\rm ho}=\sqrt{\hbar/m \omega}$ and $\hbar \omega$, respectively, with a frequency $\omega$ of a harmonic trap; thus $h_{i}= -\nabla^{2}/2 + r^{2}/2 - \Omega \hat{L}_{z}$. The atom-atom interactions are characterized by $u_{i}$ for intracomponent and $u_{12}$ for intercomponent interactions. The last term in Eq. (\ref{coupleGPfunc}) describes a coherent coupling induced by an external field, which allows atoms to change their internal state coherently \cite{Hall2,Matthews2}. Here $\omega_{\rm R}$ is the Rabi frequency and $\Delta$ is the detuning between the external field and the atomic transition; in the following we set $\Delta=0$ for simplicity. In this case, the total particle number (per unit length along the $z$ axis) $N = \int d {\bf r} \{ |\Psi_{1}|^{2} + |\Psi_{2}|^{2} \} \equiv \int d {\bf r} \rho_{\rm T}$ is conserved. By renormalizing the wave function as $\Psi_{i} \rightarrow \sqrt{N}\Psi_{i}/b_{\rm ho}$, the atom-atom interactions are written as $u_{i}=4 \pi a_{i} N$ and $u_{12}=4 \pi a_{12} N$ with the corresponding s-wave scattering lengths $a_{i}$ and $a_{12}$. 

The crucial effect of the internal coherent coupling is to induce an effective attractive interaction between the two components \cite{Ripoll3}. Putting $\Psi_{i}=|\Psi_{i}|e^{i\theta_{i}}$, we can rewrite the last two terms in Eq. (\ref{coupleGPfunc}) as $-2\omega_{\rm R}|\Psi_{1}({\bf r})| |\Psi_{2}({\bf r})| \cos\phi({\bf r})$, where $\phi ({\bf r}) \equiv \theta_{1}-\theta_{2}$ is the relative phase. To decrease it, $\phi$ must decrease and the amplitudes of the two components must increase in the overlapping region. 

\begin{figure}[btp]
\includegraphics[height=0.35\textheight]{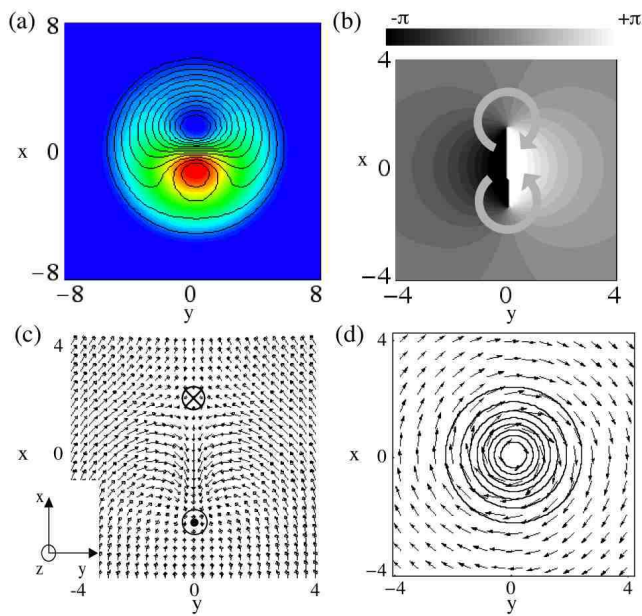}
\caption{(a) The profile of the density $|\Psi_{1}|^{2}$ (color-scale plot) and $|\Psi_{2}|^{2}$ (line contours) with $\Omega=0.15$ and $\omega_{\rm R}=0.02$. (b) The gray-scale plot of the relative phase $\phi({\bf r})=\theta_{1}-\theta_{2}$. Arrows show the direction of the circulation in the space of relative phase around the vortices. (c) The vectorial representation of the pseudospin ${\bf S}=\bar{\chi} \bm{\sigma} \chi / 2$ projected onto the $x-y$ plane. The locations of the defects are indicated by $\odot$ (with $S_{z}=+\hat{\bf z}$/2) and $\otimes$ (with $S_{z}=-\hat{\bf z}/2$). (d) The contour plot of the topological charge density $q ({\bf r})$ (the largest value at the center) and the vectorial plot of the ${\bf v}_{\rm eff}$-field.}
\label{joseffect}
\end{figure}
\begin{figure}[btp]
\includegraphics[height=0.165\textheight]{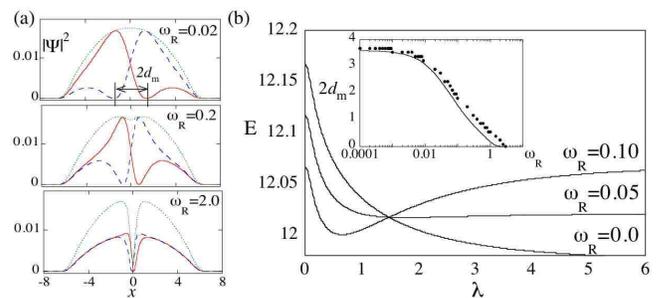}
\caption{(a) The cross sections of the condensate density along the $y=0$ line ($|\Psi_{1}|^{2}$: solid-curve, $|\Psi_{2}|^{2}$: dashed-curve, $\rho_{\rm T}$: dotted-curve). (b) The total energy as a function of $\lambda$ (the size of the meron pair) for $\Omega=0$. The inset shows the separation $2 d_{\rm m}$ between two vortices as a function of the Rabi frequency $\omega_{\rm R}$ for $\Omega=0.15$. The solid curve represents the result obtained by the variational analysis with Eqs. (\ref{nonlsigma}) and (\ref{meron}) ($d_{\rm m}=2\lambda$) and the dots the numerical result.}
\label{moledis}
\end{figure}

We first assume $u_{1}=u_{2}=u_{12}=1000$ (a more general case will be discussed later) to discuss the vortex state with the external coupling and an appropriate rotating drive to ensure stabilization of only one vortex in each component. The minimization of the energy functional (\ref{coupleGPfunc}) is done numerically by using the conjugate gradient method. Figure \ref{joseffect}(a) shows the equilibrium solution for $\omega_{\rm R}=0.02$ and $\Omega=0.15$. Each component has one off-centered vortex which shifts from the other to reduce the overlapping area [see Fig. \ref{moledis}(a)]. For finite $\omega_{\rm R}$ this nonaxisymmetric vortex state is always energetically lower than the axisymmetric one which was observed in Ref. \cite{Matthews}, in which one circulating component surrounds a nonrotating core of the other \cite{Ho}. Also, the optimized relative phase shows an unique structure as shown in Fig. \ref{joseffect}(b). Here, the central region is characterized by the configuration of a vortex-antivortex pair; the vortex cores are connected by a branch cut of the relative phase with a 2$\pi$ phase difference \cite{Son}. Thus, the two vortices attract each other, forming a bound pair, i.e., a ``vortex molecule". As $\omega_{\rm R}$ increases, the size of the pair decreases as seen in Fig. \ref{moledis}(a). Beyond $\omega_{\rm R} \simeq 3.0$ the separation vanishes, where the locations of the density nodes overlap despite the intercomponent repulsive interaction. 

An insight into the vortex molecule can be gained when we describe the two-component BECs in terms of a pseudospin \cite{Mueller,Matthews2}. By introducing a normalized spinor $\chi=[\chi_{1}({\bf r}),\chi_{2}({\bf r})]^{T}$ with $|\chi_{1}|^{2}+|\chi_{2}|^{2}=1$ and writing $\Psi_{i}=\sqrt{\rho_{\rm T}({\bf r})} \chi_{i}({\bf r})$, we can define the spin vector as ${\bf S}({\bf r})=\bar{\chi} \bm{\sigma} \chi/2$ with the Pauli matrix $\bm{\sigma}$; the modulus of the total spin is $|{\bf S}|=1/2$. The vectorial representation of the spin density (projected onto the $x-y$ plane) is shown in Fig. \ref{joseffect}(c). The spins are oriented in the $x$ direction everywhere except in the central domain-wall region where they tumble rapidly by $2\pi$. There exist two points corresponding to the locations of vortices at which ${\bf S}$ is parallel to the $z$-axis. The spin field around the singularity with ${\bf S} = +{\bf \hat{z}}/2$ (${\bf S} = -{\bf \hat{z}}/2$) has a radial (hyperbolic) distribution, having $(S_{x},S_{y}) \propto (-x,-y)$ ($\propto (x,-y)$). This texture is known as a ``radial-hyperbolic" pair of {\it merons} \cite{tyuu3}, which has been discussed not only in the study of topological defects in superfluid $^{3}$He \cite{Seppala} and a double-layer quantum Hall system \cite{Moon} but also in the semiclassical model of color confinement in QCD \cite{Steele}. 

To discuss the properties of a vortex molecule quantitatively, we rewrite Eq. (\ref{coupleGPfunc}) in terms of the pseudospin variables as
\begin{eqnarray}
E=\int d {\bf r} \biggl[ \frac{1}{2}(\nabla \sqrt{\rho_{\rm T}})^{2}+ \frac{\rho_{\rm T}}{2} (\nabla {\bf S})^{2} + \frac{\rho_{\rm T}}{2} ( {\bf v}_{\rm eff}  - {\bf \Omega} \times {\bf r} )^{2} \nonumber \\
+\frac{\rho_{\rm T}}{2} (1-\Omega^{2}) r^{2} - 2 \omega_{\rm R} \rho_{\rm T} S_{x} + \frac{\rho_{\rm T}^{2}}{2} ( c_{0} + c_{1} S_{z} + c_{2} S_{z}^{2}) \biggr], 
\label{nonlsigma}
\end{eqnarray}
where we have introduced the effective velocity field induced by spin textures
\begin{equation}
{\bf v}_{\rm eff} = \frac{\nabla \Theta}{2} +\frac{S_{z}}{(S_{x}^{2}+S_{y}^{2})} (S_{y} \nabla S_{x} - S_{x} \nabla S_{y} ) 
\end{equation}
with the total phase $\Theta=\theta_{1}+\theta_{2}$, and three coupling constants $c_{0}=(u_{1}+u_{2}+2u_{12})/4$, $c_{1}=u_{1}-u_{2}$ and $c_{2}=u_{1}+u_{2}-2u_{12}$. Equation (\ref{nonlsigma}) describes a generalized {\it nonlinear sigma model} for two-component BECs, which allows us to obtain the solutions for topological excitations such as ``skyrmions" or ``merons" \cite{Moon}. They are characterized by the topological charge density $q ({\bf r}) = \epsilon^{\mu \nu} {\bf S} \cdot (\partial_{\mu} {\bf S}) \times (\partial_{\nu} {\bf S}) / \pi = (\nabla \times {\bf v}_{\rm eff}) /2\pi$, whose spatial integral is the topological charge. Figure \ref{joseffect}(d) shows the profile of $q ({\bf r})$ together with the vector plot of the ${\bf v}_{\rm eff}$-field. Contrary to a conventional vortex, $|{\bf v}_{\rm eff}|$ vanishes at the center, reflecting a coreless vortex without a density dip in the total density. The topological charge of this solution is exactly unit. The axisymmetry of both profiles is reflected to the SU(2)-symmetry associated with $c_{1}=c_{2}=0$. Then, the structure of the meron-pair is equivalent to an {\it axisymmetric vortex state}, i.e., a skyrmion \cite{Mueller}, where both states are connected via overall spin rotation with 90 degrees \cite{tyuu}; according to our defenition of ${\bf S}$, the spin of the skyrmion points down along the $z$-axis at the center and rolls up continuously as it goes outward if the $\Psi_{1}$ component has a vortex. The internal coupling works as a transverse ``magnetic" field that aligns the spin along the $x$-axis. Therefore, turning on $\omega_{\rm R}$ makes a skyrmion split into two merons with the spin texture shown in Fig. \ref{joseffect}(c) \cite{Moon}. 

The pseudospin picture gives us an physical interpretation of vortex-molecule binding by means of the internal coupling. When $c_{1}=c_{2}=0$, the pseudospin profiles of the meron-pair can be represented by the form
\begin{equation}
S_{x}=\frac{1}{2} \frac{r^{2}-4 \lambda^{2}e^{-\alpha r^{2}}}{r^{2}+4 \lambda^{2}e^{-\alpha r^{2}}}, \hspace{4mm}
S_{y}=\frac{-2 \lambda y e^{-\alpha r^{2}/2}}{r^{2}+4 \lambda^{2}e^{-\alpha r^{2}}}, 
\label{meron}
\end{equation}
and $S_{z}=(1/2 - S_{x}^{2}-S_{y}^{2})^{1/2}$, where $\lambda$ and $\alpha$ are the variational parameters that characterize the size of the meron-pair, i.e., the molecular size $d_{\rm m}^{2} e^{\alpha d_{\rm m}^{2}}=4\lambda^{2}$ (see the top of Fig. \ref{moledis} (a) for the definition of $d_{\rm m}$). The problem becomes simpler by assuming $\alpha=0$ so that Eq. (\ref{meron}) reduces to a solution of the classical nonlinear sigma model \cite{Moon}.
Substituting {\bf S} into Eq. (\ref{nonlsigma}) and putting $\Theta=\tan^{-1} \frac{y}{x-2\lambda} + \tan^{-1} \frac{y}{x+2\lambda}$, we obtain $E=\int d {\bf r}[(\nabla \sqrt{\rho_{\rm T}})^{2}/2 + V_{\rm eff} \rho_{\rm T} + c_{0} \rho_{\rm T}^{2}/2]$ with 
\begin{equation}
V_{\rm eff} = \frac{r^{2}}{2} + \frac{r^{2}+8 \lambda^{2}}{2 (r^{2} + 4 \lambda^{2})^{2}} - \frac{\Omega r^{2}}{r^{2} + 4 \lambda^{2}} - \omega_{\rm R} \frac{r^{2} - 4 \lambda^{2}}{r^{2} + 4 \lambda^{2}}.
\end{equation}
To see the $\omega_{\rm R}$-dependence on the molecular binding, we calculate the total energy $E$ as a function of $\lambda$ in the case {\it without rotation} \cite{tyuu4}; the result is shown in Fig. \ref{moledis}(b). Here, for a given $\lambda(=d_{\rm m}/2)$ we calculate numerically $\rho_{\rm T}$ that minimizes $E$. For $\omega_{\rm R}=0$ the energy decreases monotonically with $\lambda$, which implies the repulsive interaction between two merons. Its dominant contribution comes from the second and third term of Eq. (\ref{nonlsigma}), which are the gradient energy of the pseudospin and the kinetic energy of the ${\bf v}_{\rm eff}$-field. Since the total density $\rho_{\rm T}$ is fixed as the Thomas-Fermi profile with an inverted parabola for the moderate values of $\lambda$ except $\lambda<0.15$, where the vortex core appears in $\rho_{\rm T}$ as seen in Fig. \ref{moledis}(a), the energy contribution of the other terms is almost constant. For finite $\omega_{\rm R}$, there appears an energy minimum so that the two vortices can form a bound pair; the value of $\lambda$ giving the minimum decreases with $\omega_{\rm R}$. This binding originates from a tension $T_{\rm d}$ of the domain wall of the relative phase. The binding energy can be estimated as $\sim T_{\rm d} d_{\rm m}$; for a homogeneous system $T_{\rm d} = 8 |\Psi_{1}|^{2} |\Psi_{2}|^{2} k / \rho_{\rm T}$ with the characteristic domain size $k^{-1}= (|\Psi_{1}||\Psi_{2}|/2\omega_{\rm R} \rho_{\rm T})^{1/2}$ \cite{Son}. 

The energy minimum in Fig. \ref{moledis}(b) is not ensured to be thermodynamically stable because the variational function Eq. (\ref{meron}) does not describe the motion of the center of mass of a vortex molecule. We performed the simulation of the imaginary time propagation of the Gross-Pitaevskii equation derived from Eq. (\ref{coupleGPfunc}), and found that for $\Omega=0$ the center of mass spirals out toward the edge of the condensates because of the drift instability \cite{Rokhsar}. Therefore a rotation is necessary to stabilize the vortex molecule actually. Figure \ref{moledis}(b) shows the $\omega_{\rm R}$ dependence of the molecular separation $2d_{\rm m}$ for $\Omega=0.15$. The variational analysis gives a good agreement with the numerically obtained value for $2d_{\rm m}$. 

\begin{figure}[btp]
\includegraphics[height=0.175\textheight]{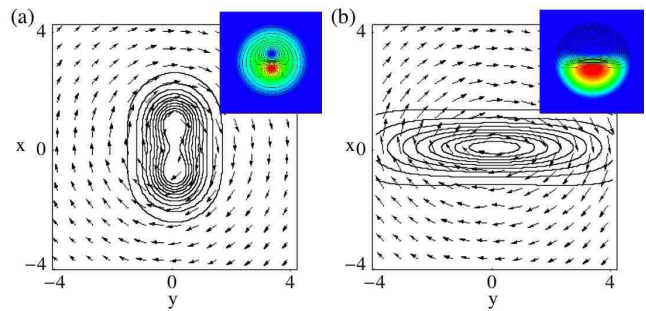}
\caption{The contour plots of the topological charge density $q ({\bf r})$ and the vectorial plots of the ${\bf v}_{\rm eff}$-field for $\Omega=0.15$, $\omega_{\rm R}=0.02$, $c_{0}=1000$ and (a) $c_{2}=20$ (antiferromagnetic), and (b) $c_{2}=-20$ (ferromagnetic). The inset shows the corresponding density profiles.}
\label{spintex}
\end{figure}
In a realistic situation, e.g., two-component BECs studied in the JILA group \cite{Hall}, the coupling constants no longer satisfy $u_{1}=u_{2}=u_{12}$. Then, there appear a longitudinal magnetic field $c_{1}$ that aligns the spin along the $z$-axis and the spin-spin interaction $c_{2}$ associated with $S_{z}$ (antiferromagnetic for $c_{2}>0$ and ferromagnetic for $c_{2}<0$ \cite{Kasamatsu}) in Eq. (\ref{nonlsigma}). The structure of the vortex molecule for $c_{2} = \pm 20$ ($c_{1}$=0) is shown in Fig. \ref{spintex}; the detailed effects of those terms on the vortex molecule will be discussed elsewhere. Then, the axisymmetry of the topological charge and the velocity field of the vortex molecule is broken. For the antiferrmagnetic (ferromagntic) case, the anisotropic vorticity is distrubuted parallel (perpendicular) to the molecular polarization. 

\begin{figure}[btp]
\includegraphics[height=0.242\textheight]{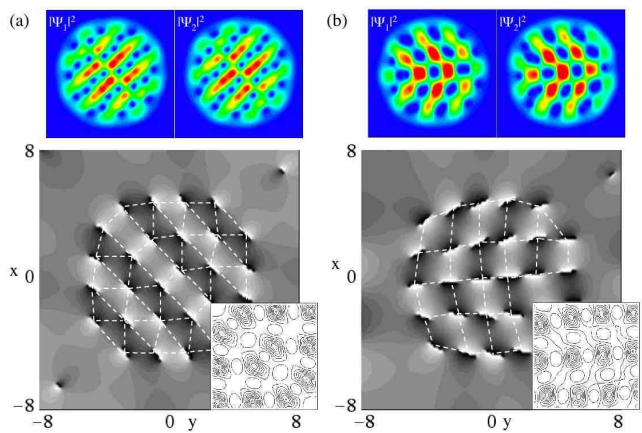}
\caption{The upper figures show the profiles of the condensate density $|\Psi_{1}|^{2}$ and $|\Psi_{2}|^{2}$ and the lower ones the relative phase for $\Omega=0.80$, $\omega_{\rm R}=0.2$ and (a) $c_{2}=20$ (antiferromagnetic), (b) $c_{2}=-20$ (ferromagnetic). The centers of mass of the molecules are linked by dashed lines. The bottom inset shows the distribution of the topological charge within [-3$<x$, $y<$3]} 
\label{spintex2}
\end{figure}

As the rotation frequency is increased, a large number of vortices are nucleated and form a lattice. When $\omega_{\rm R}$ is turned on, the vortices begin to form pairs and result in the lattice of vortex molecules. For $u_{1}=u_{2}=u_{12}$ ($c_{1}=c_{2}=0$) the vortex molecules exhibits an axisymmetric feature, yielding a square lattice of meron-pairs (equivalently, a lattice of skyrmions under the basis transformation of Ref. \cite{tyuu}). As $\omega_{\rm R}$ increases further, the vortices of one component overlap completely with those of the other, forming a triangular lattice. If $u_{1} \neq u_{2} \neq u_{12}$, however, the vortex molecule is anisotropic as shown in Fig. \ref{spintex}, so that the resultant state also exhibits a distorted lattice structure. Typical examples are shown in Fig. \ref{spintex2}, where (a) and (b) correspond to the antiferromagnetic and ferromagnetic cases, respectively. From the inset, the direction of the ordering depends strongly on the distribution of the topological charge in each vortex molecule. 

In conclusion, we have shown that an internal coherent coupling between rotating two-component BECs bring about a new effective ``vortex-molecular" field which features a (pseudo)spin texture with meron-pairs. The size of the vortex molecules can be controlled by tuning the external coupling field. This will open further interesting problems such as structural transitions and collective oscillations of the lattice of vortex molecules, and new phenomena related with double-layer quantum Hall physics. 



\begin{thebibliography}{99}
\bibitem{tyuu2}
These vortex and antivortex have circulations of the same sign in the respective phase spaces, but they are out-of-phase in the relative phase space, in contrast to a conventional vortex-antivortex pair.
\bibitem{tyuu3}
If the molecule is polarized along the $y$-axis, the texture forms a ``circular-hyperbolic" pair. Those two configurations are different just in their orientation of the molecular polarization, and can therefore continuously transform into each other.
\bibitem{tyuu}
In the SU(2) symmtric case, it is useful to rotate the basis of the spinor so that the internal coupling becomes simpler. Actually, if we rotate the spinors as $\Psi_{\pm}=(\Psi_{1} \pm \Psi_{2})/\sqrt{2}$ (the basis along with pseudo-$x$ axis), the nonaxisymmetric structure in Fig. \ref{joseffect} is transformed to the axisymmetric vortex state. Here, the vortex core of the ``+" component is filled with ``-" particles. Then, the internal coupling is just the chemical potential difference between the ``+" and ``-" components. As one increases $\lambda$ the number of ``-" particle drops, and the vortex cores evetually become empty (corresponding to the bottom of Fig. \ref{moledis}(a)).
\bibitem{tyuu4}
The rotation also produces an energy minimum without the internal coupling, which is irrelevant to the binding due to the domain wall of the relative phase.
\end{thebibliography}
\end{document}